\documentclass[amsmath,amssymb,aps,showpacs,twocolumn,prl,floatfix]{revtex4}
\usepackage{graphicx}
\usepackage{dcolumn}
\usepackage{bm}
\usepackage{color}
\usepackage{multirow}
\usepackage{times}

\newcommand{\vc}[1]{\ensuremath{\boldsymbol{#1}}}

\newcommand{\kvec}{\vc{k}}
\newcommand{\sk}{\ensuremath{\vc{s}_{\vc{k}}}}
\newcommand{\vsof}{\ensuremath{\boldsymbol{\Omega}}}
\newcommand{\vssof}{\ensuremath{\boldsymbol{\omega}}}
\newcommand{\TDP}{\ensuremath{T_{\rm DP}}}
\newcommand{\TEY}{\ensuremath{T_{\rm EY}}}
\newcommand{\sqa}{\ensuremath{\hat{s}}}

\begin{document}

\title{Strong spin-orbit fields and Dyakonov-Perel spin dephasing in supported metallic films}

\author{Nguyen H. Long} 
\email{h.nguyen@fz-juelich.de}
\affiliation{Peter Gr\"unberg Institut and Institute for Advanced Simulation, 
Forschungszentrum J\"ulich and JARA, D-52425 J\"ulich, Germany}
\author{Phivos Mavropoulos}\email{ph.mavropoulos@fz-juelich.de} 
\author{David S. G. Bauer, Bernd Zimmermann, Yuriy Mokrousov} 
\affiliation{Peter Gr\"unberg Institut and Institute for Advanced Simulation, 
Forschungszentrum J\"ulich and JARA, D-52425 J\"ulich, Germany}
\author{Stefan Bl\"ugel} 
\affiliation{Peter Gr\"unberg Institut and Institute for Advanced Simulation, 
Forschungszentrum J\"ulich and JARA, D-52425 J\"ulich, Germany}
\date{\today}

\begin{abstract}
Spin dephasing by the Dyakonov-Perel mechanism in metallic films
deposited on insulating substrates is revealed, and quantitatively
examined by means of density
functional calculations combined with a kinetic equation. 
The surface-to-substrate asymmetry, probed by the metal wave functions in
thin films, is found to produce strong spin-orbit fields and a fast
Larmor precession, giving a dominant contribution to spin decay over the Elliott-Yafet spin relaxation up to a thickness of 70 nm. 
The spin dephasing is oscillatory in time with a rapid (sub-picosecond) initial decay. However, parts of the Fermi surface act as spin traps,
causing a persistent tail signal lasting 1000 times longer than
the initial decay time. It is also found that the decay depends on the
direction of the initial spin polarization, resulting in a
spin-dephasing anisotropy of 200\% in the examined cases.
\end{abstract}

\pacs{72.25.Rb, 73.50.Bk, 72.25.Ba, 85.75.-d}

\maketitle

In spintronics experiments, spins are often excited in, or transported
through, non-magnetic metallic thin film media \cite{Jedema01}. Typical examples are Cu, Au or Pt,
used in spin-current creation or detection via the spin Hall effect \cite{Sinova04,Valenzuela06,Kimura07,Saeki08} or spin Nernst effect \cite{Cheng08,Tauber12}. 
Paramount for the spin-transport properties of a
medium is the characteristic time $T$ after which the
out-of-equilibrium spin population that was created in the medium is
lost by relaxation or dephasing \cite{Zutic04,Fabian98}.  The microscopic mechanisms leading
to spin reduction depend on the material properties, and it is
commonly accepted that the Elliott-Yafet (EY) mechanism \cite{Elliott54,Yafet61} is dominant in
metals \cite{Lubzens76,Johnson85,Gradhand09,Gradhand10}, since they show space-inversion symmetry \cite{Footnote1}.
However, any substrate on which the film is deposited breaks
the inversion symmetry; if the film is thin enough (thinner than the
electron phase relaxation length), the resulting asymmetry is felt by
the metallic states extending over the film thickness, even though the
substrate and surface potential are screened in the film interior. In
this case, as we argue in this Letter, the band structure changes throughout
the film and the Dyakonov-Perel (DP) mechanism \cite{DyakonovPerel} for spin
dephasing is activated and becomes the dominant cause of spin
reduction. The DP mechanism (that we briefly describe below) is known
to be important in III-V or II-VI semiconductors or semiconductor
heterostructures due to their inversion asymmetry \cite{Pikus84,Mower11}, but, to our
knowledge, it has been completely overlooked so far in the important case of metallic films.

Characteristic of systems with spin-orbit coupling and time reversal
symmetry but broken inversion symmetry is the lifting of
\emph{conjugation} degeneracy \cite{Yafet61} at each crystal momentum
\kvec\ and energy $E_{\kvec}$ of the band structure. The resulting
pair of states $\Psi^{\pm}_{\kvec}$ obtains energies $E^{\pm}_{\kvec}$
with the (usually small) splitting
$\hbar|\vsof_{\kvec}|=E^{+}_{\kvec}-E^{-}_{\kvec}$ depending on the
spin-orbit strength, the strength of the antisymmetric part of the potential $V_{\rm A}$, and the
overlap of the wavefunction $\Psi_{\kvec}$ with $V_{\rm A}$. The
situation is described by adding to the \kvec-dependent crystal
Hamiltonian the term
\begin{equation}
\Delta H(\kvec) = \frac{\hbar}{2}\vsof_{\kvec}\cdot \boldsymbol{\sigma}
\label{hamiltonian}
\end{equation}
where $\boldsymbol{\sigma}$ is the vector of Pauli-matrices and the vector quantity
$\vsof_{\kvec}$ is called the \emph{spin-orbit field} (SOF) \cite{Zutic04,Fabian07}. The direction of $\pm \vsof_{\kvec}$ is given by the direction of the spin expectation value of $\Psi^{\pm}_{\kvec}$. 

From this well-known theory follows the Dyakonov-Perel mechanism of
spin dephasing. In brief, one assumes that an electron wavepacket at
wave vector \kvec\ with a given spin direction $\sk$ is composed of a
superposition of $\Psi^{\pm}_{\kvec}$. Then, effectively,
$\vsof_{\kvec}$ will act on the spin as a magnetic field due to
Eq.~(\ref{hamiltonian}) and $\sk$ will precess around
$\vsof_{\kvec}$. After an average momentum lifetime $T_{\rm p}$, the
electron is scattered with a transition rate $P_{\kvec'\kvec}$ to $\kvec'$ occupying a superposition of
$\Psi^{\pm}_{\kvec'}$ (the scattering is
assumed to be energy- and spin-conserving) and precesses around
$\vsof_{\kvec'}$, etc. Since the scattering sequence is a stochastic
process, the electron spin effectively precesses around a sequence of
random axes and the information on the initial direction of $\sk$ is
finally lost after a characteristic Dyakonov-Perel time of $\TDP$. The process is
governed by a kinetic equation \cite{Fabian07}:
\begin{equation}
\frac{\partial\sk}{\partial t}=\vsof_{\kvec}\times\sk
-\sum_{\vc{k'}}P_{\vc{k}'\vc{k}}\left(\sk-\vc{s}_{\vc{k'}}\right).
\label{kineticeq}
\end{equation}
The occupation of $\Psi^{\pm}_{\kvec}$ by a single wavepacket implies
that the splitting $\hbar |\vsof_{\kvec}|$ is small: one can conceive
a wavepacket of large energy spread, but it is unlikely that after
several scattering events both $\Psi^{\pm}_{\kvec}$ will follow the
same path in \kvec-space if their energy difference is large. In this
sense, we expect that the Rashba surface states of metals, being in
many cases characterized by a large $\hbar|\vsof_{\kvec}|$ [e.g. of
the order of 100~meV at the Fermi level for Au(111) \cite{Lashell96,Henk03}], will produce
strong spin relaxation but not follow the DP mechanism.

In the present Letter we demonstrate the importance of the DP
mechanism in metallic films deposited on insulating substrates. 
We use the density-functional-based Korringa-Kohn-Rostoker (KKR) Green function method for the calculation of the band
structure and transition rates \cite{Heers12,Long13,Long14}, and Eq.~(\ref{kineticeq}) for the time
evolution of the spin expectation value. In our
calculations we explicitly assume that all scattering is caused by
self-adatom impurities, that are always present on metal surfaces; the
concept that we demonstrate, however, is valid also in the presence of
other scattering sources and can be easily quantified
if the transition rate is known. As we find, the dephasing process is
controlled by an interplay between film thickness, scattering
strength, and penetration depth of the film wavefunctions into the
insulating substrate and into the vacuum. 

In the following, we give a short description of our method of calculation of
spin-orbit fields, focussing on the basic principles and the
approximations \cite{Supp}. In a metallic film of thickness $d$ deposited on an
insulating substrate, the wavefunctions $\Psi_{\kvec}$ around the
Fermi energy probe the surface and substrate potential, $V_{\rm surf}$
at $z>d/2$ and $V_{\rm sub}$ at $z<-d/2$, only by exponentially
evanescent tails ($z=0$ defines the film mid-plane). Since by
assumption, the free-standing film shows inversion
symmetry, the antisymmetric part of the potential is just $V_{\rm A}(\vc{r})=\frac{{\rm sign}(z)}{2}[V_{\rm surf}(\vc{r})-V_{\rm sub}(-\vc{r})]$. The smallness
of the overlap $(\Psi_{\kvec},V_{\rm A}\Psi_{\kvec})\propto1/d$ allows us to
calculate $\vsof_{\kvec}$ in linear response to $V_{\rm A}$ with the
free-standing film as a reference (the linear approximation improving
at larger thicknesses).  In a second, simplifying step, the substrate
is mimicked by a constant barrier $V_0$ added to the surface potential
of the free-standing film at $z<-d/2$, yielding $V_{\rm A}=-\frac{{\rm
sign}(z)}{2}V_0\,\theta(|z|-d/2)$. The conceptual advantage of this
approximation is that one can define spin-orbit fields characteristic
of the free-standing film, where the substrate enters only via a
linear multiplicative factor $V_0$. I.e., one obtains the
linear relation
\begin{equation}
\vsof_{\kvec} = V_0 \ \vssof_{\kvec} 
\label{ssf},
\end{equation}
where the \emph{spin-orbit field susceptibility} $\vssof_{\kvec}$ was introduced. 
The value of $\vssof_{\kvec}$ can be calculated in linear response theory on the basis of the free-standing film, while the parameter
$V_0$ can be fitted at high symmetry points in the
Brillouin zone for any given susbtrate with respect to an explicit calculation of the film on
the substrate. We consider this second step well suited for a
qualitative discussion, while the exact values of $\vsof_{\kvec}$ can
deviate somewhat from this result. Quantitative improvements by taking
the full substrate potential explicitly into account are 
possible but numerically expensive and are not necessary to unravel the general phenomenon which is the motivation here.

In solving Eq.~(\ref{kineticeq}) for the spin population $\sk(t)$, we
assume that the excited electron concentration is small, so that the
scattering is practically not affected by the final state occupation,
and that the excited states are close to the Fermi level, confining in
practice the $\kvec$ values to the Fermi surface (FS). As an initial
condition at $t=0$, we choose that $\sk$ is along the positive $z$
axis, i.e.\ normal to the film surface. 
Other choices of initial
conditions, corresponding to different physical situations, are of
course possible. 
A Fermi surface integration gives us the
sought-for quantity $\langle s_z (t)\rangle = (1/n_{\rm
F})\sum_{\kvec} (\sk(t))_z$, i.e., the magnetization along
the initial axis $z$, normalized to the density of states $n_{\rm F}$
at the Fermi level. 
We assume that we are in the low concentration regime, i.e.,
$P_{\vc{k}\vc{k}'}$ scales linearly to the impurity
concentration. Even under this assumption, the solution of
Eq.~(\ref{kineticeq}) has no simple scaling properties with respect to
concentration or to $V_0$. Eq.~(\ref{kineticeq}) has to be
explicitly solved for each set of these parameters.

We also compare the DP with the EY mechanism that neglects
precession but accounts for spin-flip scattering. We employ the master
equation for the spin-dependent electron distribution function $n_{\kvec}^{\sigma}(t)$
involving spin-conserving and spin-flip transition rates
$P_{\kvec\kvec'}^{\sigma\sigma'}$ \cite{Supp},
\begin{equation}
\frac{d n_{\kvec}^{\sigma}}{dt} =
\sum_{\kvec'} [
P_{\kvec\kvec'}^{\sigma\sigma}n_{\kvec'}^\sigma +
P_{\kvec\kvec'}^{\sigma\bar{\sigma}}n_{\kvec'}^{\bar{\sigma}} -
P_{\kvec'\kvec}^{\sigma\sigma}n_{\kvec}^\sigma -
P_{\kvec'\kvec}^{\bar{\sigma}\sigma}n_{\kvec}^\sigma 
],
\label{eq:EY}
\end{equation}
where $\bar{\sigma}=\uparrow$ if
$\sigma=\downarrow$ and vice versa. A time integration
gives $\langle s_z(t)\rangle=
\frac{\hbar}{2}\frac{1}{n_F}\sum_{\kvec}(n_{\kvec}^{\uparrow}(t)-n_{\kvec}^{\downarrow}(t))$.
Contrary to Eq.~(\ref{kineticeq}), Eq.~(\ref{eq:EY}) is linear in the impurity concentration.

\begin{figure}
\includegraphics[scale=0.9,trim=10 20 1 1,clip=true]{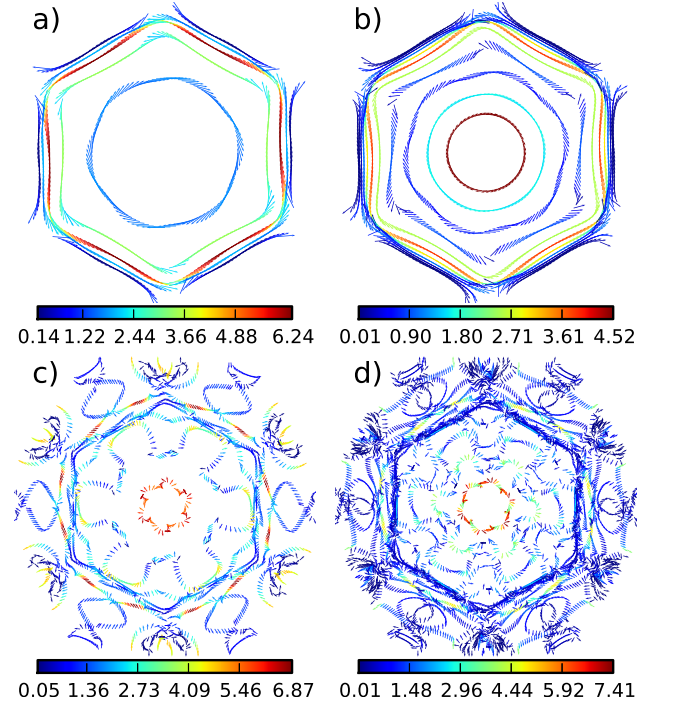}
\caption{\small (Color online) The distribution of the 
spin-orbit field susceptibility $\vssof_{\kvec}$ on the Fermi surfaces of a) 6-layer
Au(111), b) 12-layer Au(111), c) 6-layer Pt(111) and d) 12-layer
Pt(111) films. The arrows denote the projection of the direction of the spin-orbit
fields onto the surface plane. The color code denotes
the absolute value of $\hbar|\vssof_{\kvec}|\times 10^3$. The surface
states are not shown. See the supplemental material for plots with improved resolution.}
\label{combineS}
\end{figure}

As we find, the form of $\langle s_z(t) \rangle$ is rather
complicated, not having an exponential envelope. Still, we
\emph{define} the dephasing time $\TDP$ and relaxation time $\TEY$ as
the time at which $\langle s(t) \rangle = \exp(-1) \langle s
(0)\rangle $. Especially the EY mechanism causes an initial
exponential decay with decay parameter
$1/\TEY=2\sum_{\vc{k}\vc{k}'}P^{\downarrow\uparrow}_{\vc{k}\vc{k}'} /
n_{\rm F}$. In
reality, both mechanisms act simultaneously, which should be taken
into account in a full theory \cite{Boross13}. Here, however, we are after a
separation of causes, comparing the two mechanisms as if they were
independent.

We proceed with a presentation of our results. In Fig.~\ref{combineS}
we show the Fermi surface distribution of the spin-orbit field susceptibility for
Au(111) and Pt(111) 6-layer and 12-layer films (omitting surface
states).  The arrows denote the direction of the $\vssof_{\kvec}$
projected onto the surface plane while the color code shows the value
of $\hbar\left|\vssof_{\kvec}\right|\times 10^3$. We find sizeable
out-of-plane components of $\vssof_{\kvec}$ \cite{Supp}. These must
identically vanish only in the presence of in-plane inversion
symmetry, $V(x,y,z)=V(-x,-y,z)$ \cite{Footnote3}, e.g. in bcc/fcc(100) or (110)
systems. 

\begin{figure}
\includegraphics[scale=0.32,trim=60 1 25 1,clip=true,angle=270]{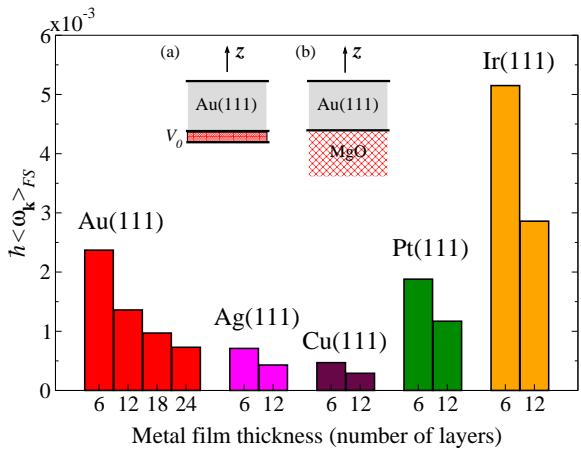}
\caption{\small (Color online) The Fermi surface average of spin-orbit
  field susceptibility $\hbar\left<|\vssof|\right>$ in various
  metallic films as a function of film thickness. The inset shows the Au(111) film a) on the model substrate $V_0$ and b) on the MgO substrate.}
\label{avaOmega}
\end{figure}

The magnitude of the spin-orbit field susceptibility shows a spread, as seen from the color code.
Averaging over the Fermi surface, we expect $\left<|\vssof|\right>
\propto 1/d$. 
Qualitatively, this behavior is indeed observed in Fig.~\ref{avaOmega}
for all tested systems: Au(111), Ag(111), Cu(111), Pt(111) and Ir(111)
\cite{Dil08}.  Comparing Au(111),
Ag(111) and Cu(111) at the same thickness, we find that stronger
spin-orbit coupling leads to larger averaged SOF, as expected from a spin-orbit phenomenon.

For an estimation of the parameter $V_0$ [Eq.~(\ref{ssf})] we
calculated self-consistently a 6 ML Au(111) film on 6 layers of
the wide-band-gap insulator MgO and found explicitly the SOF at high
symmetry points in the Brillouin zone. A fit to $V_0\vssof_{\kvec}$,
with $\vssof_{\kvec}$ calculated for a free-standing 6-atomic-layer Au(111) film,
gives us $V_0=-12.24$~eV, which we accept as characteristic
of the Au/MgO interface at all film thicknesses. 
Thus the Fermi surface average yields spin-orbit field of $\hbar\left<|\vsof|\right>=29$ meV
(for 6 layers of Pt(111) we obtain $\hbar\left<|\vsof|\right>=25$ meV).
The value decreases to 9 meV in a 24-atomic-layer Au(111) film, which is still
considerably larger than a typical value of 1 meV met in
semiconductors \cite{Zutic04}, with the consequence of a much faster spin
precession in these metallic films (a splitting of $\hbar|\vsof|
=1$~meV corresponds to a Larmor precession time of
$T_{\rm L}=4.13$~ps). Different insulating substrates are
expected to have different values of $V_0$ but in the same order of magnitude.

\begin{figure}
\includegraphics[scale=0.315,trim=60 1 25 1,clip=true,angle=270]{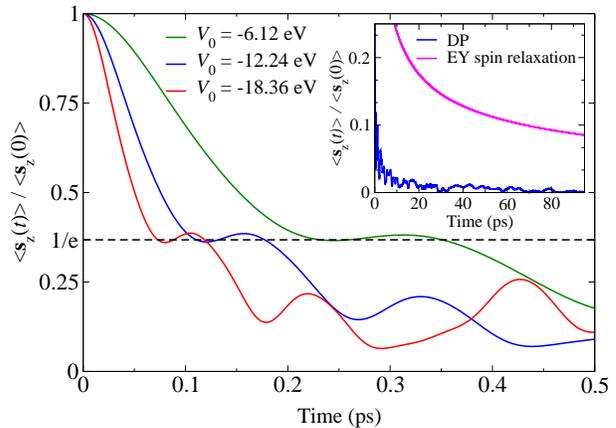}
\caption{\small (Color online) The Dyakonov-Perel time-decay of the
  total electron spin in 24-layer Au(111) films with 1\% adatom
  impurities. Three cases are considered: $V_0=-6.12$ eV, $-12.24$ eV
  and $-18.36$ eV. Inset: the case of $V_0=-12.24$ eV on a larger time scale, together with the Elliott-Yafet decay.}
\label{spinrelaxDP}
\end{figure}

Now we are ready to discuss the solution of the kinetic equation
(\ref{kineticeq}). Fig.~\ref{spinrelaxDP} shows the value of $\langle
s_z(t)\rangle/\langle s_z(0)\rangle$ for a 24-atomic-layer Au(111) film. 
The adatom
concentration is set to 1\%. The initial rapid drop gives $\TDP\sim
0.1$--$0.3$~ps. The behavior is clearly oscillatory with a
non-exponential envelope and with part of the signal ($\sim2$\% of the
initial value) persisting to times as large as 80~ps, as can be seen
from the inset of Fig.~\ref{spinrelaxDP}.
We also explored a variation of the parameter $V_0$ to $-6.12$ eV and $-18.36$
eV; the qualitative behavior is the same, but with a slower or faster
decay, respectively, due to the different precession frequency (see
Fig.~\ref{spinrelaxDP}). 

The origin of this behavior lies in the standing-wave nodal structure
that the metal wavefunctions (``quantum-well states'') exhibit in the
$z$ direction (perpendicular to the film). Depending on the band,
$|\Psi_{\vc{k}}(\vc{r})|^2$ can have either a negligible or a
significant value in the vacuum and in the substrate. In the former
case, the $\vsof_{\kvec} \sim (\Psi_{\vc{k}},V_{\rm A} \Psi_{\vc{k}})$
almost vanishes. At these $\kvec$-points, the precession term in
Eq.~(\ref{kineticeq}) is negligible even over long times. The only
path to dephasing for these spins is to be scattered away first to
some $\kvec'$ with larger SOF. But since the initial state
$\Psi_{\vc{k}}$ does not penetrate in the vacuum, the overlap with the
adatom potential is also small, keeping the scattering rate low. These
particular parts of the FS act in a sense as \emph{spin traps}. Since
the nodal structure of the quantum-well states depends primarily on
the metallic film and not on the substrate or on the adatom, the spin
traps are a property of the pristine film. However, in the presence of
burried impurities instead of adatoms, or of phonons at elevated
temperatures, the scattering rate will not be negligible and the spins
will be scattered away from the traps at a higher rate.  Additionally, 
the precession can freeze if $\vsof_{\kvec}$ and $\sk$ are collinear.
Since this condition is met in part of the FS (not shown) of Au(111) or Pt(111), it is part of
the reason of the slow decay of $s_z$ in these systems. We should note
that the existence and effectiveness of spin traps is material and
thickness dependent.  Fig.~\ref{spinrelaxDP} (inset) also shows $\langle
s_z(t)\rangle/\langle s_z(0)\rangle$ by the EY mechanism (Eq.~\ref{eq:EY}). Evidently,
the EY spin relaxation is also affected by the spin traps, producing
persisting tails, and is not exponential at large times. It is clear,
however, that the DP mechanism dominates the decay process.

\begin{table}
\begin{tabular}{cccccccc}
 & \multicolumn{3}{c}{$\TDP$ (ps)} & \multicolumn{2}{c}{$\TEY$
 (ps)} & \multicolumn{2}{c}{$T_{\rm p}$ (ps)}\\ \hline
 $V_0$ (eV)  & $-$6.12 & $-$12.24 & $-$18.36 & ws & wos & ws & wos \\ \hline
6-layer, 1\% imp. & 0.054  & 0.027  & 0.018  & 5.84 & 16.06 & 0.10 & 0.51 \\ 
6-layer, 5\% imp. & 0.059  & 0.029  & 0.018  & 1.16 & 3.20 & 0.02 & 0.10 \\ 
24-layer, 1\% imp. & 0.22 & 0.11  & 0.072  & 1.27 & 47.65 & 0.56 & 1.06 \\ 
24-layer, 5\% imp. & 0.31 & 0.17 & 0.083  & 0.25 & 9.53 & 0.11 & 0.21 \\ 
\hline
\end{tabular}
\caption{Spin-depasing time induced by the Dyakonov-Perel
mechanism, \TDP, in comparison to the spin-relaxation time \TEY\
induced by the Elliott-Yafet mechanism and the
momentum-relaxation time $T_{\rm p}$  in
6-layer Au(111) and 24-layer Au(111) films with Au adatoms as
scatterers. The adatom concentration is taken with respect to
full surface coverage. \TEY\ and $T_{\rm p}$ are given with (ws) and
without (wos) the surface states taken into account.}
\label{tauDP}
\end{table}

Table~\ref{tauDP} summarizes the values of $\TDP$ for 6-atomic-layer
and 24-atomic-layer Au(111) with 1\% and 5\% of self-adatoms and for
different substrate-potential values $V_0$. The same Table also shows
the calculated EY relaxation time, including values with (``ws'') and
without (``wos'') the surface states taken into account. (The latter
serves for comparing the DP and EY mechanisms acting on the same set
of states.) As is qualitatively expected \cite{Zutic04}, \TDP\
increases and \TEY\ is reduced with increasing defect concentration.
It is striking that \TDP\ is in all cases much lower than \TEY. The
reason for this is basically the very high value of the spin-orbit
fields causing a Larmor precession period $T_{\rm L}$ that is smaller
than the momentum relaxation time $T_{\rm p}$.  E.g., for 24
layers of Au(111), $\left<|\vsof|\right>=9$~meV giving $T_{\rm L}=0.46$~ps,
to be compared to $T_{\rm p}\sim 1$~ps at 1\% adatom
concentration.
In this regime, according to Zuti\'c et al. \cite{Zutic04}, $\TDP$ is estimated as inverse of the SOF spread $\Delta\Omega$. 
For 24 layers of Au(111), $\TDP\sim 1/\Delta\Omega=67$~fs, which is in the same
order as our first-principles value of $\TDP=110$~fs.  Only at much larger
thicknesses does the average precession period exceed a few ps,
allowing the EY mechanism to dominate \cite{Footnote4}. 
Assuming that
$\left<|\vsof|\right>\propto 1/d$, we can estimate that the two
mechanisms will have a comparable contribution at thicknesses of
approximate 270 layers (70 nm). The DP effect, however, could be washed
out at elevated temperatures, if the electron phase relaxation length
becomes smaller than the film thickness so that electrons in the film
interior cannot probe the substrate-surface asymmetry (different to semiconductors, the asymmetry potential does not penetrate deep into the metallic film interior due to metallic screening). In cases of weak spin-orbit coupling, e.g. Li,
Na, Mg or Al, both the SOF and the spin-mixing of states are reduced
proportionally to the spin-orbit strength. As a result the DP
mechanism is still expected to dominate at small thicknesses, however,
it is expected to enter the motional narrwowing regime because of
the slower Larmor precession \cite{Zutic04}.

Finally, one expects an anisotropy of the DP dephasing time with
respect to the initial condition, e.g., \TDP\ will be different for
$\vc{s}_{\kvec}(t=0)$ perpendicular to the film compared to its being
in the film plane. This type of anisotropy has been reported
previously e.g. in semiconductor heterostructures or in graphene
\cite{Fratini13} and originates from a different microscopic mechanism
compared to the one reported for inversion-symmetric metals \cite{Zimmermann13,Mokrousov13,Long13-2}. Using the symbol $\sqa$ to denote the
direction of the initial spin polarization, we have a dependence
$\TDP(\sqa)$ and we may define the anisotropy as the relative
difference
\begin{equation}
\mathcal{A_{\rm DP}} = \frac{\max_{\sqa}\TDP(\sqa) -\min_{\sqa}\TDP(\sqa) }{\min_{\sqa}\TDP(\sqa)}. 
\label{eq:ansisotropy}
\end{equation}
For the films studied here, $\mathcal{A_{\rm DP}}$ reaches values as
large as 200\% in a 24-atomic-layer Au(111) film. The spin traps are persistent
irrespective of the initial condition, as they originate in
regions of very small SOF, irrespective of the \sqa.

In conclusion, we find that the potential asymmetry introduced in supported metallic films by the substrate will create strong spin-orbit
fields, activating the DP mechanism of spin dephasing. 
This effect is present in spite of the strong charge-screening in metals, since the surface and substrate are probed by the itinerant metal wavefunctions even if the potential perturbation vanishes in the film interior. 
We predict that the DP mechanism can dominate over the EY mechanism for
thicknesses as large as 200-300 atomic layers, after which the interface induced SOF become very small, falling off inversely proportional with thickness of the film. 
In case of scattering only by adatoms, certain parts of the Fermi surface with low SOF acquire also low scattering rates, acting as spin traps and allowing the spin signal to persist over long times. 
We propose that an experiment to verify our predictions can be based on laser-pulse, pump-probe experiments (probing the Faraday or Kerr rotation) with fs temporal resolution in films of varying thickness.

We would like to thank Gustav Bihlmayer, Jaroslav Fabian and Swantje Heers for fruitful discussions.  This work was financially supported by Deutsche Forschungsgemeinschaft projects MO 1731/3-1 and SPP 1538 SpinCaT, and the HGF-YIG NG-513 project of the Helmholtz Gemeinschaft.  We acknowledge computing time on the supercomputers JUQEEN and JUROPA at J\"ulich Supercomputing Center and JARA-HPC Compute cluster of RWTH Aachen University.


\begin{thebibliography}{99}
\bibitem{Footnote3} This is because the in-plane reflection operator $R_{xy}$ transforms $\Psi_{\vc{k}}$ to $\Psi_{-\vc{k}}$ which is also the time-reverse state; thus the $z$-component of the spin, $s_z$, must flip by time-reversal. At the same time, $R_{xy} \sigma_z R_{xy}^{-1}=+\sigma_z$, i.e., $s_z$ must be conserved.  This leads to $s_z=0$, i.e., also $(\vsof_{\vc{k}})_z=0$.
\bibitem{Footnote4} $T_{\rm p}$ is, to a first approximation, independent of the film thickness: the scattering matrix element off impurities scales as $\propto 1/d$, but the number of available final states increase as $\propto d$.
\end{thebibliography}
\end{document}